\documentclass[12pt,dvips]{article}
\usepackage{rotating}
\usepackage{epsfig}
\usepackage{color}
\usepackage{cite}
\usepackage[small]{caption}
\usepackage{multirow}

\newcommand{\psl}{\mathbf{p} \hspace{-0.5 em}/}
\newcommand{\Esl}{E \hspace{-0.5 em}/}

\begin{document}

\def\thefootnote{\fnsymbol{footnote}}

\begin{flushright}
  IFT-UAM/CSIC-11-46
\end{flushright}

\begin{center}
  {\bf {\LARGE
      Reconstructing 
      the heavy resonance \\[2mm]
      at hadron colliders
    }}
\end{center}

\medskip

\begin{center}{\large
    Chan~Beom~Park
  }
\end{center}

\begin{center}
  {\em Instituto de F\'isica Te\'orica UAM/CSIC,\\
    Nicol\'as Cabrera 13-15, Universidad Aut\'onoma de Madrid,\\
    Cantoblanco, 28049 Madrid, Spain}\\[0.2cm]
\end{center}

\bigskip

\centerline{\bf ABSTRACT}
  \medskip\noindent
  We investigate the method for constructing the invariant
  mass using the $M_{T2}$-assisted on-shell (MAOS) approximation
  to the invisible particle momenta in the cascade decays of a new particle resonance
  produced at hadron colliders.
  We note that the MAOS reconstruction can be defined in several
  different ways, while keeping the efficiency of approximation at a similar level,
  and one of them provides a unique solution for each event.
  It is shown that the invariant mass distribution
  constructed with the MAOS momenta exhibits a peak
  at the heavy resonance mass, regardless of the chosen
  MAOS scheme and 
the detailed mass spectrum of the particles in the cascade.
  We stress that the MAOS invariant mass can be used
  as a clean signal of new
  particle resonance produced at the Large Hadron Collider, 
as well as a model-independent
  method to measure the masses of new particles involved in the event.

\newpage

\section{Introduction}
\label{sec:intro}

The naturalness principle predicts that there is new physics
beyond the standard model (SM) at TeV scale, which is the scale of the
Large Hadron Collider (LHC) experiment running now at CERN.
It has also been claimed that one of the most important ingredients for
the new physics model is the existence of a viable dark matter
candidate, which is usually a weakly interacting massive particle
(WIMP).
In large classes of the new physics scenarios, the WIMP is often
stabilized by a discrete parity, which consequently yields a generic
collider signature with the missing particles always in pairs.
Well-known examples such as supersymmetric (SUSY), little Higgs,
and extra-dimensional models include a WIMP, which is the
lightest new particle stabilized by a $Z_2$ parity~\cite{Nilles:1983ge,ArkaniHamed:2001nc,Appelquist:2000nn}.
The typical collider signature of the new physics model with
conserved $Z_2$ parity is the production of a new particle pair,
decaying to some visible SM particles (multiple leptons and/or jets)
{\em plus} invisible WIMPs in the final state.

To uncover the underlying physics, one needs the information of the
particle properties such as mass and spin, which
can be revealed from the reconstruction of the signal events.
However, in hadron colliders, it is generically impossible
because the center-of-mass frame of the partonic
collision is unknown and there are two invisible WIMPs
in each event.
Still, it has been proposed that the new particle masses
might be determined by various kinematic variables such as the
end points of the invariant mass distributions, transverse
variables, and the techniques using the on-shell mass
constraints in various new physics processes with the pair-produced
new particles~\cite{Barr:2010zj}.

In this paper, we consider a new physics
event,\footnote{The event topology is the same as the ``antler''
  diagram, which was studied in \cite{Han:2009ss}. However, here we
  assume that $m_X$ is unknown {\em a priori}.}
\begin{eqnarray}
  p\,\,p(\bar{p}) \to X + U \to Y_1\,\,Y_2 + U \to
  V_1(p)\chi_1(k)\,\,V_2(q)\chi_2(l) + U(u) ,
  \label{eq:eventType}
\end{eqnarray}
where $X$,
$Y_i$, and $\chi_i$ ($i$ = 1, 2) are new particles with
{\em a priori} unknown mass,
and $U$ stands for visible particles not
associated with the decay of $X$ such as jets from the
initial state radiation.
We assume that $X$ has even $Z_2$ parity, while $Y_i$ and
$\chi_i$ have odd $Z_2$ parity.
Consequently, $X$ can be resonantly produced and decay
into a parity-odd particle pair.
$V_i$'s denote visible particles and $\chi_i$'s invisible ones,
which yield the missing transverse
momentum of the event,
\begin{eqnarray}
  {\bf \psl}_T = -{\bf p}_T - {\bf q}_T - {\bf u}_T
  = {\bf k}_T + {\bf l}_T .
  \label{eq:missing_ET}
\end{eqnarray}
A similar event topology
has recently received a lot of attention.
It describes the dilepton channel of the top-pair produced
in the $s$-channel mediation of the color sextet
bosons~\cite{Berger:2010fy}, or axigluon~\cite{Sehgal:1987wi} at
hadron colliders. In this case, the eight unknown components of two
neutrino four-momenta can be solved from the six on-shell mass
conditions on the top quarks, $W$ bosons, and neutrinos in addition
to the two constraints from the missing transverse momentum
measurements,
up to four possible solutions, as well as two combinations
due to the charge ambiguity on $b$
quarks~\cite{Sonnenschein:2006ud}.
In fact, the event topology (\ref{eq:eventType}) is one of the typical
signatures of the new physics model with the WIMP stabilized by $Z_2$
parity, for instance,
the heavy neutral SUSY Higgs boson ($H/A$) decaying to a pair of
neutralinos ($\tilde\chi_i^0$)~\cite{Baer:1992kd,Bisset:2007mi}
or the $n=2$ Kaluza-Klein (KK) $Z$-boson ($Z^{(2)}$) decaying to a
pair of $n=1$ KK leptons ($l^{(1)}$)~\cite{Cheng:2002iz},
producing the final state of several visible SM particles $+$
WIMPs.
The masses of the new particles involved in the decay chain are
generically unknown, 
and the number of decays in the chain might be too short
to constrain the unknown masses and invisible momenta
in many new physics scenarios unlike the aforementioned
top-pair process.

It has been claimed in the literature that it might be possible to
measure the particle masses by constructing the transverse
mass variables even if there are several invisible particles
in the final state~\cite{Barr:2009mx,Tovey:2010de,Barr:2011he}.
However, we note that the end-point position of
the transverse mass distribution depends on the existence of
kinematic configurations, which might be forbidden
in some models.
On the other hand,
it has been recently found that the $M_{T2}$-assisted on-shell
(MAOS) method~\cite{Cho:2008tj},
which was introduced to approximate the invisible
particle momenta in the physics processes
with conserved $Z_2$ parity,
can be used to measure the SM Higgs boson mass in the dileptonic $WW$
process~\cite{Choi:2009hn}.
We show that the MAOS method can be adopted to construct the
invariant mass in the case of the event topology (\ref{eq:eventType}),
thus enabling one to measure the resonance mass
in a model-independent way.

This paper is organized as follows.
In Sec.~\ref{sec:m_t}, we discuss the features of the
transverse mass variable for the full system, focusing on
the behavior of its end point.
Then, in Sec.~\ref{sec:m_t2}, we describe the definition
and the types of solutions for the $M_{T2}$ variable,
which is an integral part of constructing the MAOS momenta,
as well as known to be useful for measuring the masses of the
intermediate on-shell states and the WIMP.
For each type of the $M_{T2}$ solution, the MAOS momenta
have distinctive features, which lead to proposing various
schemes of the MAOS reconstruction,
as discussed in Sec.~\ref{sec:maosMass},
where the construction of the
invariant mass using the MAOS momenta is also described.
The comparison between the MAOS schemes and the features
of the MAOS invariant mass are shown by performing a Monte Carlo
(MC) study in Sec.~\ref{sec:monte}.
We summarize our conclusions in Sec.~\ref{sec:concl}.

\section{Transverse mass of the full system: $M_T$}
\label{sec:m_t}

In this section, we discuss some features of the transverse
mass for the event topology (\ref{eq:eventType}).
The impossibility of constructing the invariant mass of the decay
system due to the existence of invisible particles in the final state
leads to the proposal of a transverse mass variable, which does not
use the unknown longitudinal component of the
momenta~\cite{Smith:1983aa}.
The transverse mass for the event type (\ref{eq:eventType}) can be
written as
\begin{eqnarray}
  M_T^2\left(Y_1Y_2\right) = m_{V_1V_2}^2 + m_{\chi_1\chi_2}^2
  + 2\left(\sqrt{|{\bf p}_T^{V_1V_2}|^2 + m_{V_1V_2}^2}
    \sqrt{|{\bf p}_T^{\chi_1\chi_2}|^2 + m_{\chi_1\chi_2}^2}
    - {\bf p}_T^{V_1V_2} \cdot {\bf p}_T^{\chi_1\chi_2}\right) ,
\end{eqnarray}
where $m_\alpha$ and ${\bf p}_T^\alpha$ denote the invariant mass and
the transverse momentum, respectively of
$\alpha=V_1V_2,\,\chi_1\chi_2$.
It is then obvious that $M_T(Y_1Y_2)$
is bounded from above by $m_X$, thus making the determination of
$m_X$ possible if $M_T(Y_1Y_2)$ is correctly constructed event
by event.
It is, however, generally impossible to determine
$m_{\chi_1\chi_2}$ event by event, even though the lower bound might be
deduced from the knowledge of the event topology and $m_{\chi_i}$
values in some cases, e.g., the SM Higgs boson which decays
into two leptonically decaying $W$ bosons
($h\to WW\to l\nu l^\prime\nu^\prime$).
If the lower bound of $m_{\chi_1\chi_2}$ is determined,
it must be always true that
\begin{eqnarray}
  M_T(Y_1Y_2)|_{m_{\chi_1\chi_2} = m_{\chi_1\chi_2}^{\rm min}}
  \leq
  M_T(Y_1Y_2)|_{m_{\chi_1\chi_2} = m_{\chi_1\chi_2}^{\rm true}}
  \leq m_X ,
\end{eqnarray}
since $M_T(Y_1Y_2)$ is a monotonically increasing function of
$m_{\chi_1\chi_2}$.
This fact unties us from ignorance of the
event-by-event values of $m_{\chi_1\chi_2}$, and consequently allows us
to construct the transverse mass, which can be used to
extract the information of $m_X$~\cite{Barr:2009mx,Tovey:2010de}.
For the sake of discussion from now on, we focus on the symmetric
decay chains, i.e. $m_{Y_1}=m_{Y_2}=m_Y$
and $m_{\chi_1}=m_{\chi_2}=m_\chi$.
However, we notice that the arguments below can be generalized to the
case of asymmetric decay chains.

If $m_\chi$ is unknown and there is no viable theoretical assumption
on the lower bound of $m_{\chi\chi}$, the best choice will be
\begin{eqnarray}
  M_T^2(0) \equiv M_T^2(Y_1Y_2)|_{m_{\chi_1\chi_2}=0}
  = m_{V_1V_2}^2 + 2|\psl_T| \sqrt{|{\bf p}_T+{\bf q}_T|^2+ m_{V_1V2}^2}
  - 2({\bf p}_T+{\bf q}_T)\cdot\psl_T ,
\end{eqnarray}
then $M_T^{\rm max}(0) = m_X$ if and only if 
$m_{\chi_1\chi_2}^{\rm min} = 0$.
The maximum of $M_T(0)$ occurs 
when $\chi_i$'s are moving parallel to each other in the $X$ rest
frame,
and the invariant mass of the visible particles is minimized.
Consequently, 
the analytic expression of the $M_T^{\rm max}(0)$ is 
obtained as~\cite{Tovey:2010de}
\begin{eqnarray}
  M_T^{\rm max}(0) = \frac{m_X}{2m_Y^2}\left(
    \Lambda + \sqrt{\Lambda^2 - 4m_Y^2 \left(m_V^{\rm min}\right)^2}
  \right) ,
\end{eqnarray}
where $\Lambda \equiv m_Y^2 - m_\chi^2 + (m_V^{\rm min})^2$ and
$m_V^{\rm min} \equiv \min\{m_{V_1},\,m_{V_2}\}$ for all the events.
If $m_V^{\rm min}=0$,
\begin{eqnarray}
  M_T^{\rm max}(0) = m_X \left(1-\frac{m_\chi^2}{m_Y^2}\right),
\end{eqnarray}
which shows that the $M_T(0)$ distribution can never reach $m_X$ if
$\chi_i$ is massive. The $M_T(0)$ variable was found to be useful
for measuring the SM Higgs boson mass in dileptonic $WW$ decay
mode~\cite{Barr:2009mx}.

On the other hand, in many new physics scenarios, $m_\chi$ is
likely to be determined by the other kinematic variables such as the
$M_{T2}$ kink, which will be described in Sec.~\ref{sec:m_t2},
or possibly the combinations of the invariant mass
end points from various other decay processes.
In such cases, the more plausible choice of the
$m_{\chi_1\chi_2}^{\rm min}$ value will be
$2m_\chi$ ($=m_{\chi_1}+m_{\chi_2}$) as
\begin{eqnarray}
  M_T^2(2m_\chi) = m_{V_1V_2}^2 + 4m_\chi^2
  + 2\sqrt{|{\bf p}_T + {\bf q}_T|^2 + m_{V_1V_2}^2}
  \sqrt{|\psl_T|^2 + 4m_\chi^2}
  - 2\left({\bf p}_T + {\bf q}_T\right)\cdot\psl_T .
\end{eqnarray}
But still, the knowledge of $m_\chi$ does not guarantee the saturation
of the bound on $m_X$.
If there is no kinematic configuration in which
$m_{\chi_1\chi_2} = 2m_\chi$ is achieved,
the $M_T(2m_\chi)$ distribution will not reach $m_X$, because
\begin{eqnarray}
  M_T(Y_1Y_2)|_{m_{\chi\chi}=2m_\chi} <
  M_T(Y_1Y_2)|_{m_{\chi\chi}=m_{\chi\chi}^{\rm min}} \leq
  M_T(Y_1Y_2)|_{m_{\chi\chi}=m_{\chi\chi}^{\rm true}} \leq
  m_X .
\end{eqnarray}
The condition for the existence of the kinematic configuration
saturating the bound of $M_T(2m_\chi)$ up to $m_X$
depends on the mass pattern of the involved particles in the decay
channel, and has been derived in \cite{Tovey:2010de}:
\begin{eqnarray}
  m_X \leq \frac{m_Y^2 + m_\chi^2 - \left(m_V^{\rm
        min}\right)^2}{m_\chi} .
  \label{eq:conditionMax}
\end{eqnarray}
One can see that it may be impossible to satisfy the above condition
if $X$ is too heavy, compared with the mass scale of its decay
products.
In case the condition (\ref{eq:conditionMax}) is not satisfied,
$m_X$ is above the upper bound of $M_T(2m_\chi)$, which depends
on the possible range of $m_{V_1V_2}$:
\begin{eqnarray}
  M_T^{\rm max}(2m_\chi) &=&
  2m_\chi + m_{V_1V_2}^{\rm max}
  \nonumber\\
  &=&
  2m_\chi + \frac{m_X}{2m_Y^2}\left(
    \Lambda + \sqrt{1-\frac{4m_Y^2}{m_X^2}}
    \sqrt{\Lambda^2
      -4m_Y^2 (m_V^{\rm min})^2}
  \right) .
\end{eqnarray}

\begin{figure}[t!]
  \begin{center}
    \epsfig{figure=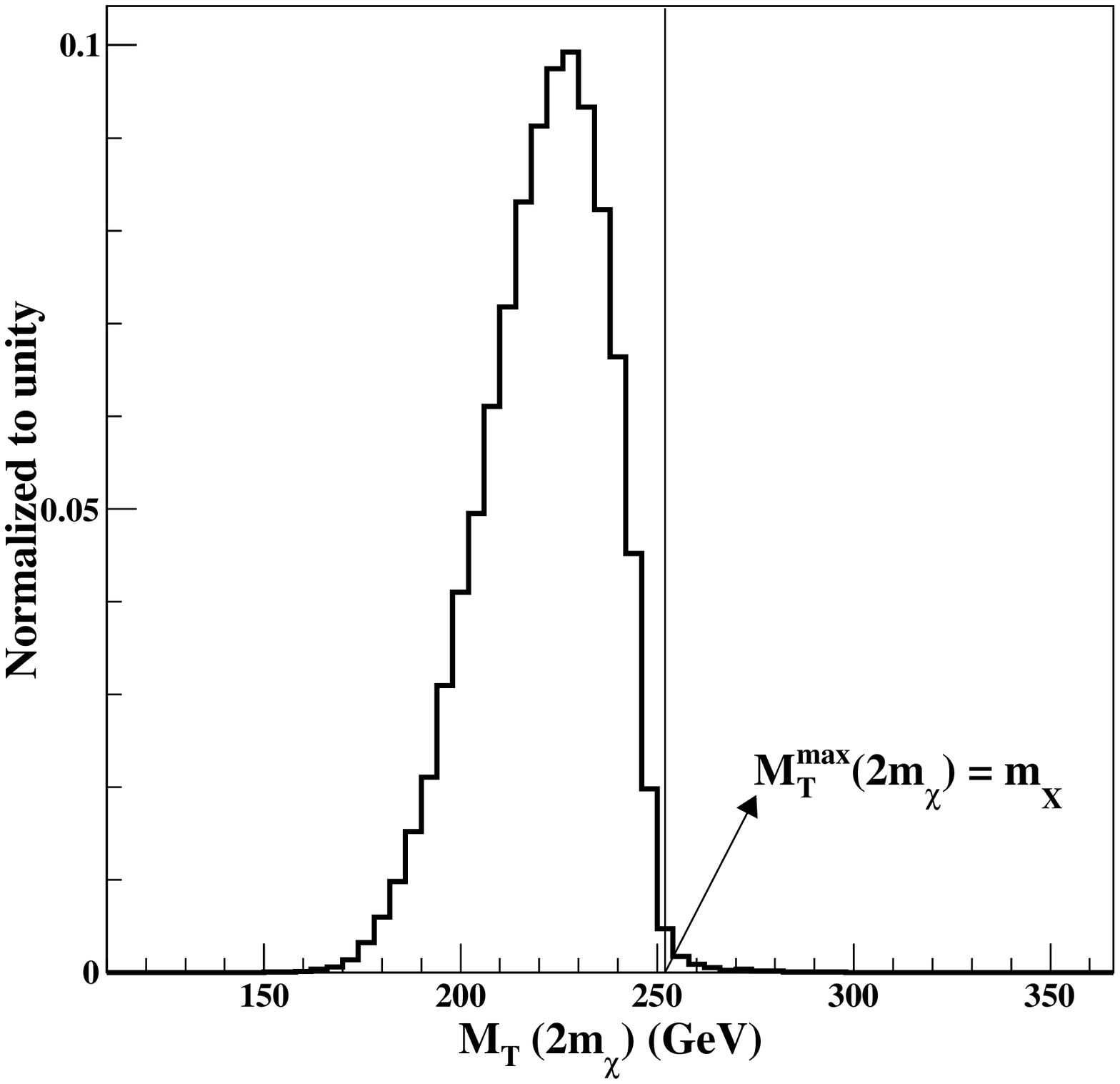,width=8cm}
    \epsfig{figure=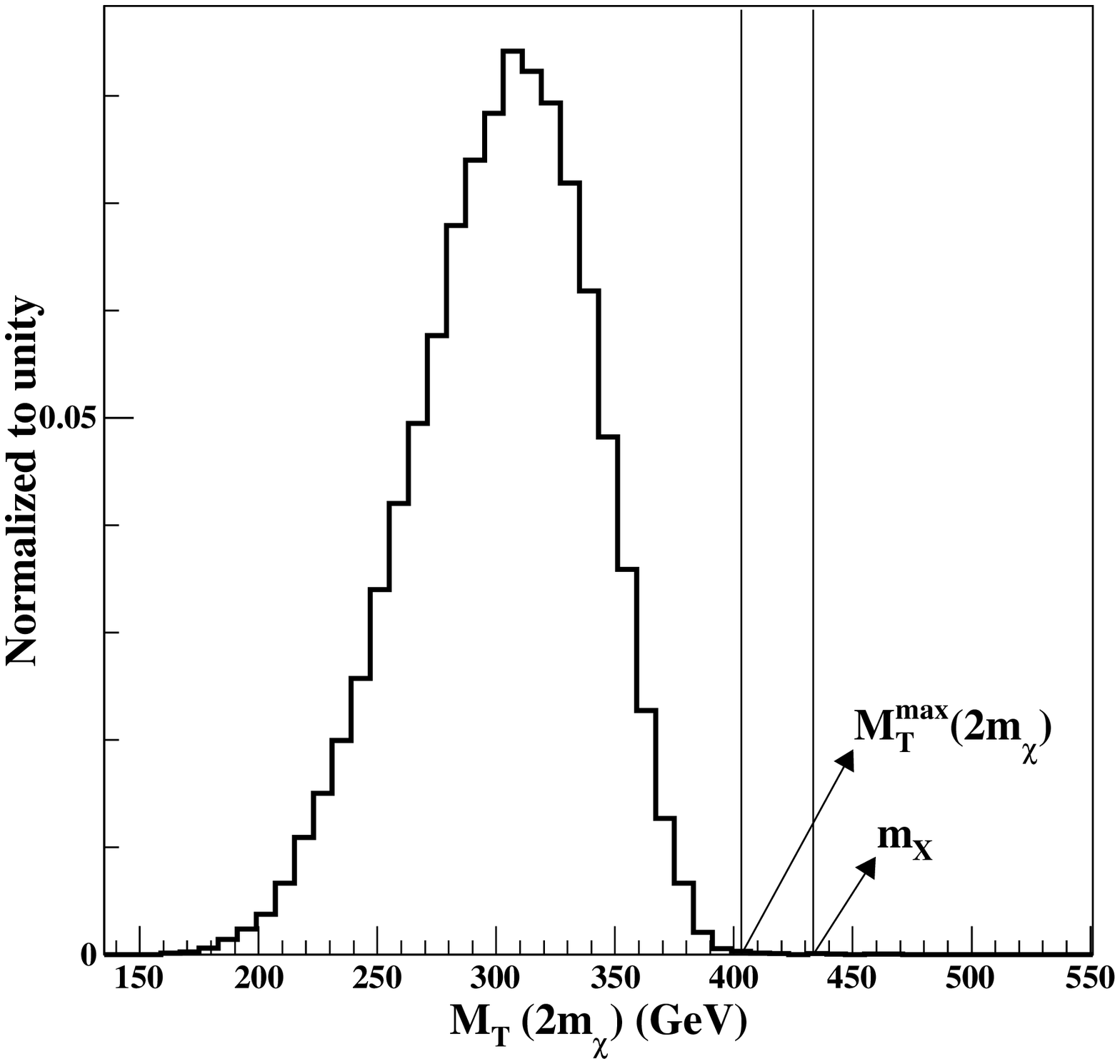,width=8cm}
  \end{center}
  \caption{Comparison plot of the  $M_{T}(2m_\chi)$ distributions of  the
    SUSY process $H/A\to \tilde\chi_2^0\tilde\chi_2^0 \to
    l^+ l^- \tilde\chi_1^0 l^+ l^- \tilde\chi_1^0$.
    The mass parameters are $(m_{\tilde\chi_2^0},\,m_{\tilde\chi_1^0})
    \simeq$ (110, 61) GeV for both plots, and
    $m_{H/A} =$ 252 (433) GeV for the left (right) panel.
   See Sec.~\ref{sec:monte} for a
   detailed description of the model and its simulation.
  }
  \label{fig:mt2mchi_distribution}
\end{figure}
In order to check the properties of $M_T(2m_\chi)$ discussed above,
we have generated MC event samples of heavy neutral Higgs bosons
($H/A$) for two SUSY benchmark points.
In this model, the heavy Higgs boson decays to a pair of
next-to-lightest neutralinos ($\tilde\chi_2^0$), producing
the final state of four charged leptons and a lightest
neutralino ($\tilde\chi_1^0$) pair via a three-body process,
$\tilde\chi_2^0 \to l^+l^- \tilde\chi_1^0$.
The detailed description of the chosen model is given in
Sec.~\ref{sec:monte}.
In Fig.~\ref{fig:mt2mchi_distribution}, we exhibit
the $M_T(2m_\chi)$ distribution for benchmark points
with relatively light (left panel) and heavy (right panel)
$H/A$.
One finds that the condition (\ref{eq:conditionMax})
is satisfied in the case of a relatively light $H/A$ scenario,
such that the $M_T^{\rm max}(2m_\chi)$ corresponds to $m_{H/A}$.
However, it is not satisfied in the heavy $H/A$ case,
thus $M_T^{\rm max}(2m_\chi)$ is lower than $m_{H/A}$.
This observation shows that the $M_T$ method depends highly
on the underlying model, i.e. the mass splitting of the on-shell
states involved in the event.
Hence, one needs accurate knowledge of $m_Y$ and possibly
the range of $m_V$ as well as $m_\chi$ to determine $m_X$.

\section{Alternative transverse mass: $M_{T2}$}
\label{sec:m_t2}

In a situation where there are two invisible particles in the event,
one may exploit the event variable $M_{T2}$, which was proposed
to measure the particle masses in an event topology such as $Y_1\,Y_2\to
V_1(p)\chi_1(k)\,V_2(q)\chi_2(l)$, which is,
for instance, the typical collider event of
pair-produced SUSY particles~\cite{Lester:1999tx}.
Not only is the $M_{T2}$ useful for measuring the new
particle masses even in the event topology (\ref{eq:eventType}),
but it is also an integral part of the definition
of the MAOS momenta, which will be described in the next section.
$M_{T2}$ can also be defined
in the asymmetric decay event with $m_{Y_1} \neq m_{Y_2}$ and
$m_{\chi_1} \neq m_{\chi_2}$~\cite{Barr:2009jv}.
However, here we consider only the event type with the symmetric
decay chains as in the previous section.
The $M_{T2}$ variable is defined as
\begin{eqnarray}
  M_{T2} \equiv \min_{\tilde{\bf k}_T+\tilde{\bf l}_T=\psl_T}
  \left[\max
    \left\{ M_T^{(1)}({\bf p}_T,\,\tilde{\bf k}_T,\,\tilde m_\chi),\,
      M_T^{(2)}({\bf q}_T,\,\tilde{\bf l}_T,\,\tilde m_\chi)
    \right\}\right] ,
  \label{eq:MT2_def}
\end{eqnarray}
where $M_T^{(1)}$ and $M_T^{(2)}$ are the transverse masses of the $Y_1$
and $Y_2$ systems, respectively,
\begin{eqnarray}
  \left(M_T^{(1)}\right)^2 &=& m_{V_1}^2 + \tilde m_\chi^2 +
  2\sqrt{|{\bf p}_T|^2 + m_{V_1}^2}\sqrt{|\tilde{\bf k}_T|^2
    + \tilde m_\chi^2}
  - 2{\bf p}_T\cdot\tilde{\bf k}_T ,
  \nonumber\\
  \left(M_T^{(2)}\right)^2 &=& m_{V_2}^2 + \tilde m_\chi^2 +
  2\sqrt{|{\bf q}_T|^2 + m_{V_2}^2}\sqrt{|\tilde{\bf l}_T|^2
    + \tilde m_\chi^2}
  - 2{\bf q}_T\cdot\tilde{\bf l}_T .
  \label{eq:MT_1_2}
\end{eqnarray}
Here, $\tilde m_\chi$, $\tilde{\bf k}_T$, and $\tilde{\bf l}_T$
are input trial values of the mass and transverse momenta of
the invisible particles.
They are {\em hypothesized} values that parametrize our
ignorance of the allocation of missing transverse
momenta into the invisible particle momenta in each event,
in addition to the true value of $m_\chi$.
The momentum constraint in the minimization
automatically guarantees that the sum of the hypothesized momenta is
equal to that of the true momenta, which can be determined event by
event.
When the input trial mass $\tilde m_\chi$ equal to the true
mass $m_\chi$,
\begin{eqnarray}
  M_{T2}^{\rm max} \left(\tilde m_\chi=m_\chi\right) = m_Y ,
\end{eqnarray}
provided that the parent particles $Y_i$ are on shell.

The $M_{T2}$ solution for the hypothesized
momenta can be classified into two configurations, {\em unbalanced} and
{\em balanced}.
The value of $M_{T2}$ is given by $M_T^{(1)} = M_T^{(2)}$ in the
balanced configuration, while the unbalanced $M_{T2}$ solution
is achieved when the condition for the balanced configuration,
which will be shown shortly, is not valid.
\begin{figure}[t!]
  \begin{center}
    \epsfig{figure=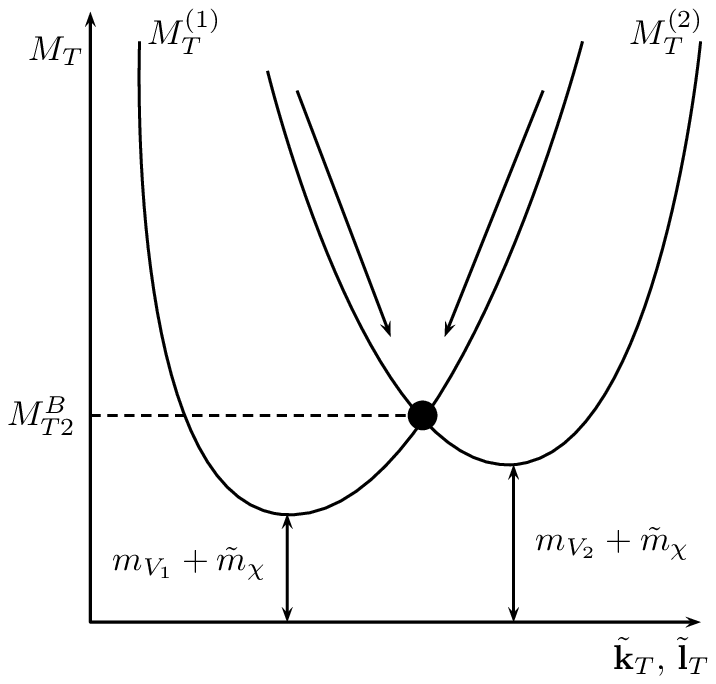,width=8cm}
    \epsfig{figure=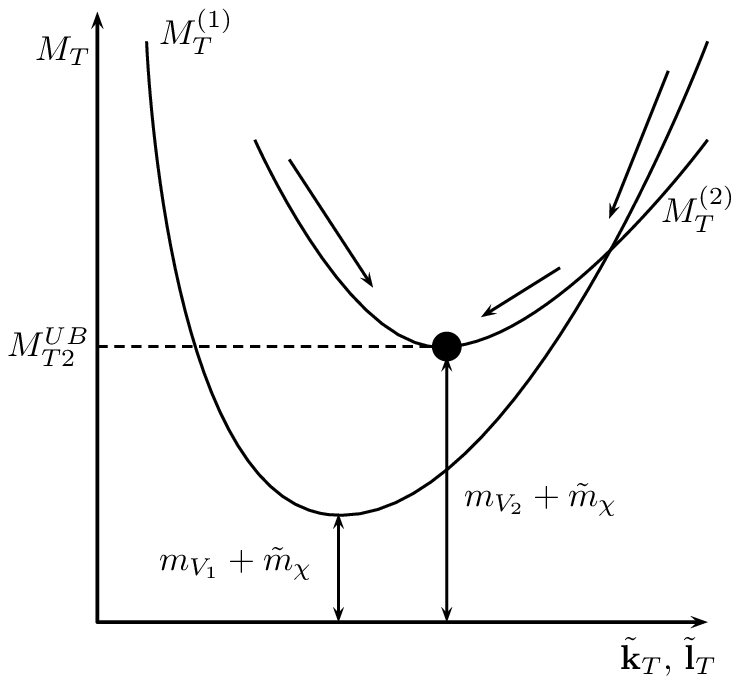,width=8cm}
  \end{center}
  \caption{A schematic picture of the balanced (left panel) and the
  unbalanced (right panel) $M_{T2}$ solutions.}
  \label{fig:mt2solution}
\end{figure}
The type of the $M_{T2}$ solution can be deduced 
from the value of $M_T^{(i)}$
at the hypothesized momenta which give the global minimum of
$M_T^{(j)}$ ($i \neq j$).
The global minimum of $M_T^{(1)}$ ($M_T^{(2)}$) corresponds to
the stationary point for given values of $m_{V_1}$ ($m_{V_2}$)
and $\tilde m_\chi$,
\begin{eqnarray}
  \left(M_T^{(1)}\right)_{\rm min} = m_{V_1} + \tilde m_\chi ,\quad
  \left(M_T^{(2)}\right)_{\rm min} = m_{V_2} + \tilde m_\chi ,
\end{eqnarray}
when $\tilde{\bf k}_T = (\tilde m_\chi/m_{V_1}) {\bf p}_T$ and
$\tilde{\bf l}_T = (\tilde m_\chi/m_{V_2}) {\bf q}_T$,
respectively.
The $M_{T2}$ value of the event 
will be obtained from the balanced configuration when 
\begin{eqnarray}
  M_T^{(i)} \geq \left(M_T^{(j)}\right)_{\rm min} \quad (i\neq j)
  \label{eq:balanced_condition}
\end{eqnarray}
is satisfied at the stationary points. [See the left panel of
Fig.~\ref{fig:mt2solution} for illustration.]
For events in which the condition (\ref{eq:balanced_condition})
is not satisfied, the $M_{T2}$ is given by the larger
value between global minima of $M_T^{(i)}$ at the stationary
points,
\begin{eqnarray}
  M_{T2}^{UB} = \max\left\{\left(M_T^{(1)}\right)_{\rm min},\,
    \left(M_T^{(2)}\right)_{\rm min}\right\} = \left (
    \begin{array}{cc}
      m_{V_1}+\tilde m_\chi & \mbox{if}\,\,\,M_T^{(1)}>M_T^{(2)} ,
      \\
      m_{V_2}+\tilde m_\chi & \mbox{if}\,\,\,M_T^{(1)}<M_T^{(2)} .
    \end{array}
  \right .
  \label{eq:unbalanced_mt2}
\end{eqnarray}
On the other hand, the solution for the balanced configuration needs
some nontrivial consideration.
The analytic expression of $M_{T2}$ was first derived
in \cite{Lester:2007fq}, then further simplified in \cite{Cho:2007qv},
by considering the event types with vanishing upstream
transverse momentum (UTM), 
i.e. ${\bf u}_T=0$ in (\ref{eq:missing_ET}).\footnote{
  Recently, the analytic expression of $M_{T2}$ has been derived
  for some special kinematic configurations.
  See Ref.~\cite{Lester:2011nj}.
}
It is given by
\begin{eqnarray}
  \left(M_{T2}^{B}\right)^2 &=& \left(M_T^{(1)}\right)^2
  = \left(M_T^{(2)}\right)^2\nonumber\\
  &=& \tilde m_\chi^2 + A_T + \sqrt{
    \left(1 + \frac{4\tilde m_\chi^2}{2A_T - m_{V_1}^2 - m_{V_2}^2}\right)
    \left(A_T^2 - m_{V_1}^2 m_{V_2}^2\right)} ,
  \label{eq:m_t2_bal}
\end{eqnarray}
where
\begin{eqnarray}
  A_T\equiv \sqrt{|{\bf p}_T|^2 + m_{V_1}^2}\sqrt{|{\bf q}_T|^2 + m_{V_2}^2}
  + {\bf p}_T\cdot{\bf q}_T ,
\end{eqnarray}
which is invariant under the back-to-back transverse boost.
This quantity is closely related to the $M_{CT}$ in
\cite{Tovey:2008ui}, which has been further generalized to the 
$M_{CT2}$ in the case of two invisible particles in the
event~\cite{Cho:2009ve}.

The investigation of the $M_{T2}$ from the
above expressions showed that the end-point position of the $M_{T2}$
distribution, $M_{T2}^{\rm max}$, as a function of the 
$\tilde m_\chi$ has a kink structure at $\tilde m_\chi = m_\chi$
if the invariant mass of the visible particles
is not fixed, but has a certain range 
in each decay chain~\cite{Cho:2007qv}.
It was also noticed that the kink structure of $M_{T2}^{\rm
  max}(\tilde m_\chi)$
will appear when there is a sizable amount of UTM in the events~\cite{Gripaios:2007is}.
The $M_{T2}$-kink method makes it possible to measure $m_\chi$ and $m_Y$
simultaneously even if the decay chain is not long enough to 
constrain all the unknowns in the event.
This observation suggests that the $M_{T2}$-kink method might also be
very useful to measure $m_Y$ and $m_\chi$ in the event topology
(\ref{eq:eventType}), specifically, if $V_i$ is a set of more
than two visible particles and/or the full system of $Y_1Y_2$
is boosted by the UTM.

If $\tilde m_\chi=m_{V_1}=m_{V_2}=0$, Eq. (\ref{eq:m_t2_bal})
becomes even simpler as
\begin{eqnarray}
  \left(M_{T2}\right)^2|_{\tilde m_\chi=0} = 2A_T
  =2\left(|{\bf p}_T||{\bf q}_T| + {\bf p}_T\cdot{\bf q}_T\right) .
\end{eqnarray}
Interestingly, the authors of \cite{Choi:2009hn} have recently
claimed that the $M_{T2}$ method can
be applied to measure the mass of the SM Higgs boson
in dileptonic $WW$ mode,
which is also the event type (\ref{eq:eventType}),
and used as an event selection cut to efficiently suppress the
background processes.

\section{Construction of the invariant mass using MAOS momenta}
\label{sec:maosMass}

After discussing the transverse mass variables,
we deal with the construction method of the invariant mass.
In Ref.~\cite{Cho:2008tj},
the authors introduced
a systematic approximation to the invisible momenta by combining the
$M_{T2}$ solution with the on-shell relations
for generic events like $Y_1\,Y_2\to
V_1(p)\chi_1(k)\,V_2(q)\chi_2(l)$.
The MAOS momenta was originally proposed to
determine the spin of new particles produced at hadron
colliders~\cite{Cho:2008tj,Nojiri:2011qn,Eboli:2011bq},
but it was later realized that it might also be very useful
for the mass measurement~\cite{Choi:2009hn,Asano:2010ii}.

The definition of MAOS momenta is composed of two parts,
the transverse and the longitudinal components.
The transverse components of the invisible momenta are set by
the trial momenta which give the value of $M_{T2}$,
\begin{eqnarray}
  {\bf k}_T^{\rm maos} = \tilde{\bf k}_T,\quad
  {\bf l}_T^{\rm maos} = \tilde{\bf l}_T,
\end{eqnarray}
where $\tilde{\bf k}_T$ and $\tilde{\bf l}_T$ are determined
once we minimize $\max\{M_T^{(1)},\,M_T^{(2)}\}$ in
(\ref{eq:MT2_def}),
among all possible trial $\tilde{\bf k}_T$ and
$\tilde{\bf l}_T$ satisfying $\tilde{\bf k}_T+\tilde{\bf l}_T
={\bf \psl}_T$.
The longitudinal and energy components are then calculated by the
on-shell relations for both $\chi_i$ and $Y_i$,
\begin{eqnarray}
    &&\left(k^{\rm maos}\right)^2 = \tilde m_{\chi_1}^2,\quad\quad\,\,\,\,
    \left(l^{\rm maos}\right)^2 = \tilde m_{\chi_2}^2,
    \label{eq:on-shell_cond_chi}
    \\
    &&\left(p+k^{\rm maos}\right)^2 = \tilde m_{Y_1}^2,\quad
    \left(q+l^{\rm maos}\right)^2 = \tilde m_{Y_2}^2,
    \label{eq:on-shell_cond_Y1}
\end{eqnarray}
where $\tilde m_{\chi_i}$ and $\tilde m_{Y_i}$ ($i$ = 1, 2)
are input trial masses of the invisible and the parent particles,
respectively.
The longitudinal components of the MAOS momenta are then given by
\begin{eqnarray}
  k_L^{\rm maos} &=& \frac{1}{E_T^2(p)}
  \left[{\cal A}p_L \pm
    \sqrt{p_L^2+E_T^2(p)}\sqrt{{\cal A}^2-E_T^2(p)E_T^2(k)}
  \right],\nonumber\\
  l_L^{\rm maos} &=& \frac{1}{E_T^2(q)}
  \left[{\cal B}q_L \pm
    \sqrt{q_L^2+E_T^2(q)}\sqrt{{\cal B}^2-E_T^2(q)E_T^2(l)}
  \right],
  \label{eq:maos_longitudinal}
\end{eqnarray}
where
\begin{eqnarray}
  \begin{array}{ll}
    {\cal A} = \frac{1}{2}\left(\tilde m_{Y_1}^2 - \tilde m_{\chi_1}^2 - m_{V_1}^2\right)
    + {\bf p}_T\cdot{\bf k}_T^{\rm maos},&
    {\cal B} = \frac{1}{2}\left(\tilde m_{Y_2}^2 - \tilde m_{\chi_2}^2 - m_{V_2}^2\right)
    + {\bf q}_T\cdot{\bf l}_T^{\rm maos},\\
    E_T(p) = \sqrt{|{\bf p}_T|^2+m_{V_1}^2},&
    E_T(q) = \sqrt{|{\bf q}_T|^2+m_{V_2}^2},\\
    E_T(k) = \sqrt{|{\bf k}_T^{\rm maos}|^2+\tilde m_{\chi_1}^2},&
    E_T(l) = \sqrt{|{\bf l}_T^{\rm maos}|^2+\tilde m_{\chi_2}^2} .
  \end{array}
\end{eqnarray}
From the above expressions,
it is obvious that both $k_L^{\rm maos}$ and $l_L^{\rm maos}$ are real
if and only if
\begin{eqnarray}
  |{\cal A}|\geq E_T(p) E_T(k),\quad
  |{\cal B}|\geq E_T(q) E_T(l) ,\nonumber
\end{eqnarray}
or equivalently
\begin{eqnarray}
  \tilde m_{Y_1} \geq M_T^{(1)} ,\quad
  \tilde m_{Y_2} \geq M_T^{(2)} ,
\end{eqnarray}
where $M_T^{(1)}$ and $M_T^{(2)}$ are the transverse masses
defined in (\ref{eq:MT_1_2}) for
$\tilde{\bf k}_T={\bf k}_T^{\rm maos}$,
$\tilde{\bf l}_T={\bf l}_T^{\rm maos}$ and the trial masses $\tilde m_{\chi_i}$.
If the decay chain is symmetric, i.e. $\tilde m_{\chi_1}=
\tilde m_{\chi_2}=\tilde m_\chi$ and $\tilde m_{Y_1}=\tilde
m_{Y_2}=\tilde m_Y$,
the above condition reduces to
\begin{eqnarray}
  \tilde m_Y \geq \max\left\{
    M_T^{(1)},\,M_T^{(2)}
  \right\},
  \label{eq:real_condition}
\end{eqnarray}
so that the MAOS momenta are always real if the value of
$M_{T2}^{\rm max} (\tilde m_\chi)$ is chosen as the trial mass of the
parent particle for a given $\tilde m_\chi$.
Note that the true mass values $m_Y$ and $m_\chi$ 
automatically satisfy the condition (\ref{eq:real_condition})
because $M_{T2}^{\rm max}(\tilde m_\chi = m_\chi)=m_Y$.
For the end-point events of the balanced $M_{T2}$,
one has $M_T^{(1)} = M_T^{(2)} = \tilde m_Y$,
and thus $k^{\rm maos}$ and $l^{\rm maos}$ correspond to the
unique solution of the constraints (\ref{eq:missing_ET}),
(\ref{eq:on-shell_cond_chi}), and (\ref{eq:on-shell_cond_Y1}).
In this case, since the true invisible momenta also satisfy
the same conditions, they must be equal to the unique
solution when the true mass values $m_\chi$ and $m_Y$
are inserted in Eqs.~(\ref{eq:on-shell_cond_chi}) and
(\ref{eq:on-shell_cond_Y1}),
\begin{eqnarray}
  k^{\rm maos} = k^{\rm true} ,\quad
  l^{\rm maos} = l^{\rm true} .
\end{eqnarray}
On the other hand, this is not true for the
end-point events of the unbalanced $M_{T2}$
because $M_T^{(1)} \neq M_T^{(2)}$, which means that
only one side of the MAOS momenta corresponds to the true
invisible momenta.
This argument indicates that the accuracy of the MAOS momenta
can be controlled by imposing a suitable $M_{T2}$ cut, which selects
the subset of events near the $M_{T2}$ end point.
Even if $m_\chi$ or $m_Y$ were poorly measured, it was shown that
the MAOS momenta would provide a good approximation to the true
invisible momenta by setting $\tilde m_\chi = 0$ and $\tilde m_Y =
M_{T2}^{\rm max}\left(\tilde m_\chi = 0\right)$~\cite{Cho:2008tj}.
One inevitable problem in this definition is that
there is generically a four-fold ambiguity
on the longitudinal and energy components as can be seen
in Eqs.~(\ref{eq:maos_longitudinal}).
In the absence of any extra constraints or viable assumptions
on the kinematic structure of the event, it is clear that
there should be no preference of one solution to the others.
We label this scheme of obtaining the solution of the invisible
momenta as the first kind of MAOS or MAOS1 to distinguish it
from the other schemes in what follows.

In addition to the ambiguity of the solutions, 
a serious problem will arise 
when one or both parent particles $Y_i$ are off shell.
The on-shell conditions (\ref{eq:on-shell_cond_Y1})
can be adopted only if the parent particles are on shell, such
that the mass values are fixed for all the events.
To make the MAOS method applicable to the situation where
the on-shell conditions are not valid,
one may consider modifying the on-shell relations.
One possible scheme is to substitute the
event variable $M_{T2}$, instead of the fixed value $\tilde m_Y$,
by
\begin{eqnarray}
  \left(p+k^{\rm maos}\right)^2 = M_{T2}^2 ,\quad
  \left(q+l^{\rm maos}\right)^2 = M_{T2}^2 .
  \label{eq:modified_scheme}
\end{eqnarray}
Then, for the events of the balanced $M_{T2}$, one has
$M_T^{(1)} = M_T^{(2)} = M_{T2}$, and thus $k_L^{\rm maos}$
and $l_L^{\rm maos}$ become unique (MAOS2).
This is the scheme which was used to measure the SM Higgs boson mass
in \cite{Choi:2009hn}, for both $m_h \geq 2m_W$
and $m_h<2m_W$ cases.\footnote{
  Note that there is only balanced configuration for the
  SM Higgs decay events, $h\to WW\to l\nu l^\prime\nu^\prime$.}
On the other hand, 
for the events of the unbalanced $M_{T2}$, i.e.
$M_T^{(1)} \neq M_T^{(2)}$,
a two-fold ambiguity still remains on one
side of the longitudinal components. 
%

The only possible way to obtain the unique solution of the
longitudinal and energy components for all the events is to take
\begin{eqnarray}
  \left(p+k^{\rm maos}\right)^2 = \left(M_T^{(1)}\right)^2 ,\quad
  \left(q+l^{\rm maos}\right)^2 = \left(M_T^{(2)}\right)^2 .
  \label{eq:unique_scheme}
\end{eqnarray}
Adopting the above new scheme, the longitudinal components  
are given by
\begin{eqnarray}
  k_L^{\rm maos} = \frac{E_T(k)}{E_T(p)} p_L ,\quad
  l_L^{\rm moas} = \frac{E_T(l)}{E_T(q)} q_L ,
\end{eqnarray}
which are uniquely defined in the events for both
balanced and unbalanced $M_{T2}$ (MAOS3).
Note that if the true mass $m_\chi$ is chosen as the input,
the MAOS momenta will be equal to the true invisible momenta
for the end-point events of the balanced $M_{T2}$ because
$M_T^{(1)} = M_T^{(2)} = M_{T2}^{\rm max}(m_\chi) = m_Y$ corresponds
to the true parent particle mass.
Although it is not true for the events of the unbalanced $M_{T2}$,
one side of the MAOS momenta ($k^{\rm maos}$ if
$M_T^{(1)} = M_{T2}^{\rm max} = m_Y > M_T^{(2)}$) still gives the true
momenta if the end-point events of $M_{T2}$ were selected.
Accordingly, one finds that the accuracy of the MAOS momenta
can be still controlled by the $M_{T2}$ cut, which might also be
useful for suppressing the SM backgrounds
in the search for a new physics signal at the LHC~\cite{Barr:2009wu}.

Applying the MAOS method to the event topology (\ref{eq:eventType}),
it is possible to construct the invariant mass of the $Y_1Y_2$ system,
\begin{eqnarray}
  \left(p+q+k^{\rm maos}+l^{\rm maos}\right)^2
  \equiv \left(m_X^{\rm maos}\right)^2 .
  \label{eq:maosMass}
\end{eqnarray}
Then, the successful reconstruction of the invisible momenta 
by the MAOS method ensures that 
the MAOS invariant mass ($m_X^{\rm maos}$)
distribution has a peak
at the true mass $m_X$, which will also be confirmed by the numerical
simulation in the next section.
The distinctive feature of the MAOS method is that it is less
model dependent than transverse mass variables
since the latter highly depends on the mass gaps in the model.
Even when the resonance particle $X$ is too heavy,
the peak position of the MAOS invariant mass distribution
corresponds to $m_X$, which might be beyond the transverse mass
distribution.
The requisites for constructing the MAOS invariant mass
are the assumption of the event topology (\ref{eq:eventType})
and the knowledge of $m_\chi$, which are also essential
for constructing $M_{T}(2m_\chi)$ in Sec.~\ref{sec:m_t}.
We also note that the MAOS invariant mass is distinguished by
the peak structure, and it is generally less
vulnerable to the background and momentum smearing effects.
The clear peak structure of the MAOS invariant mass
can be viewed as the smoking-gun signal
of the heavy resonance produced at hadron colliders.

\section{Monte Carlo study: heavy SUSY Higgs bosons}
\label{sec:monte}

In this section, we illustrate the discussion
of the previous section by performing a MC study
for the decay signal of heavy SUSY Higgs bosons $H/A$
into a pair of next-to-lightest neutralino, $\tilde\chi_2^0$,
followed by the decay $\tilde\chi_2^0\to l^+l^-\tilde\chi_1^0$
($l = e,\,\mu$).
This process results in a four-lepton
plus missing transverse energy final state,
\begin{eqnarray}
  H/A \to \tilde\chi_2^0 \tilde\chi_2^0
  \to l^+l^-l^+l^- + \Esl_T
  \label{eq:heavy_higgs}
\end{eqnarray}
As a specific example, we examined two benchmark points
in the minimal supergravity scenario chosen in \cite{Bisset:2007mi}.
The superparticle mass spectrum has been calculated with
{\tt SOFTSUSY}~\cite{Allanach:2001kg}, and is given by
\begin{itemize}
  \item point A: $m_{H/A} =$ 252 GeV, $m_{\tilde\chi_2^0}=$ 110 GeV,
    $m_{\tilde\chi_1^0} =$ 61 GeV; and
  \item point B: $m_{H/A} =$ 433 GeV, $m_{\tilde\chi_2^0}=$ 112 GeV,
    $m_{\tilde\chi_1^0} =$ 62 GeV.
\end{itemize}
We used {\tt PYTHIA} 6.4 to generate the MC events in the
LHC beam condition with the proton-proton center-of-mass energy
of 14 TeV~\cite{Sjostrand:2006za}.
For the simple illustrative study,
we have not considered the detector effects such as the
momentum smearing effect 
and the identification efficiency of the leptons.
The integrated luminosity is assumed to be large enough
to measure the particle spectra.
The main background process from the SM will be $ZZ^\ast/\gamma^\ast$
with four leptons in the final state.
This can be eliminated by requiring large missing energy and
imposing a $Z$-veto cut, which rejects events of a
dilepton pair with the invariant mass near $m_Z$.
The dominant source of the background in the SUSY process is
the production of leptons from the decays of neutralinos
and charginos, produced by squarks and gluinos.
In this case, however, the leptons are produced in
association with jets, so that a jet-veto cut should be
imposed to suppress this type of background.
The direct production of a neutralino pair via the Drell-Yan
processes could be challenging because it has the same
final state.
See Ref.~\cite{Bisset:2007mi} and Chap.~11 of \cite{Ball:2007zza}
for a detailed study of the signature including the
backgrounds at detector-level.
Here, we do not consider the background effect and
the event selection cuts except the $M_{T2}$ cut.

In the above benchmark points, $\tilde\chi_2^0$ decays
to $\tilde\chi_1^0$ and two charged leptons via three-body process with
an off-shell intermediate $Z$ boson or slepton ($m_{\tilde l_R} =$
141 and 406 GeV for points A and B, respectively),
such that
$0 \leq m_{l^+ l^-} \leq m_{\tilde\chi_2^0}-m_{\tilde\chi_1^0}$.
The main difference between the benchmark points is the
mass of $H/A$, which is relatively light in point A and
heavier in point B.
One can easily find that the condition (\ref{eq:conditionMax})
is not satisfied in point B, 
whereas it is satisfied in point A.
This results in the position of $M_T^{\rm max}(2m_\chi)$
being lower than $m_{H/A}$, as can be seen in the right panel of
Fig.~\ref{fig:mt2mchi_distribution}.

\begin{figure}[t!]
  \begin{center}
    \epsfig{figure=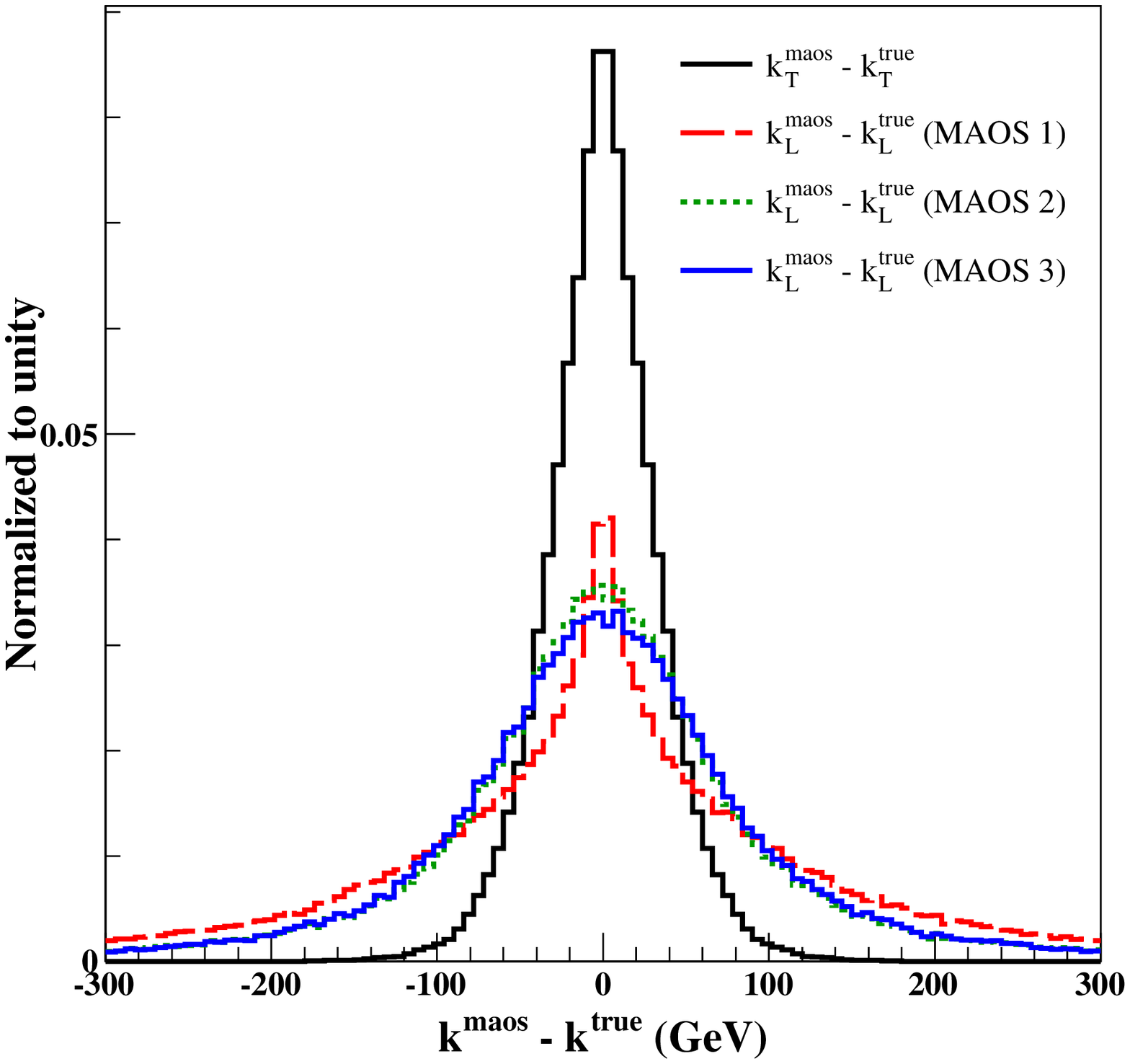,width=8cm}
    \epsfig{figure=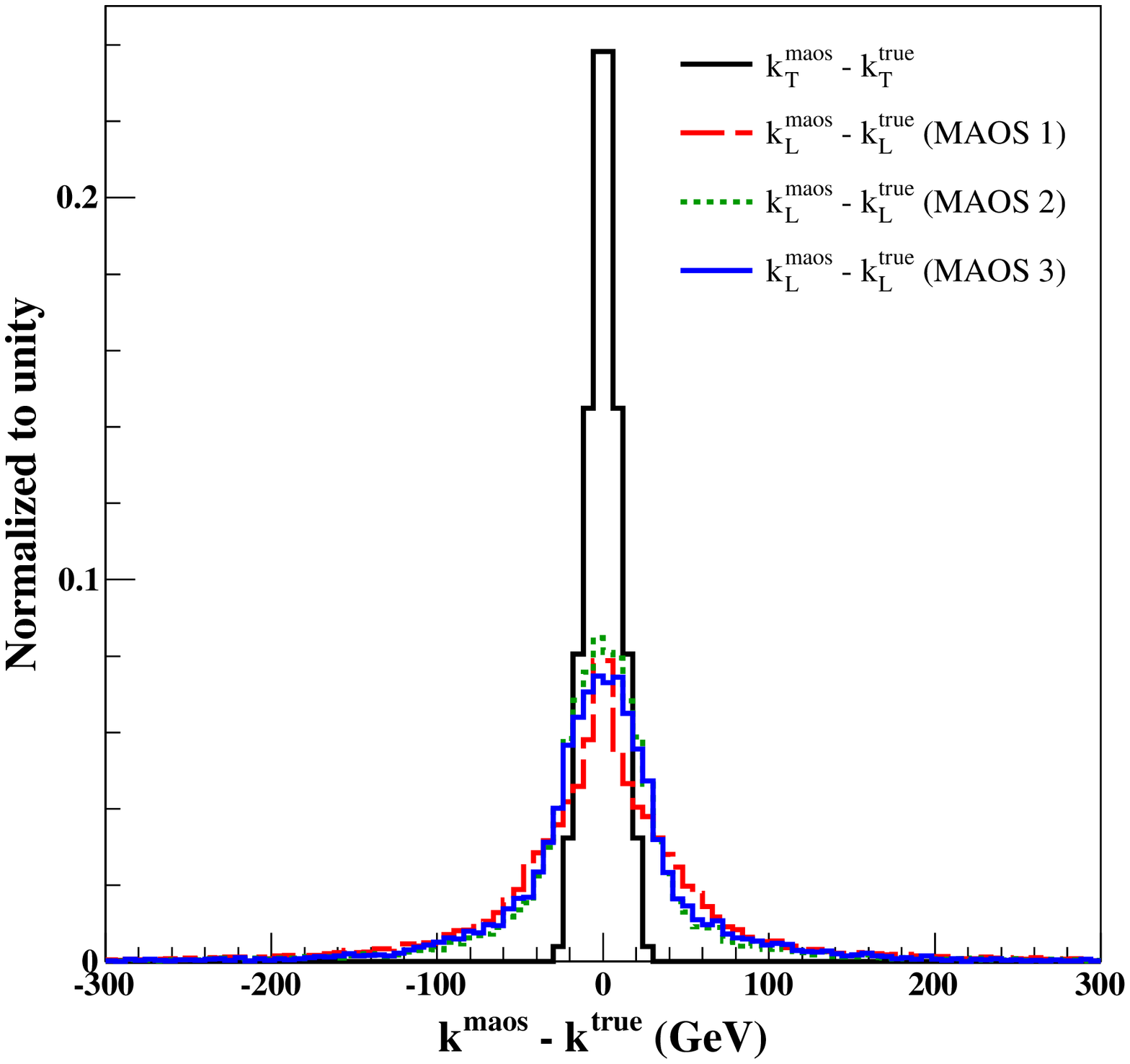,width=8cm}
  \end{center}
  \caption{The distributions of ${\bf k}_T^{\rm maos} - {\bf k}_T^{\rm
      true}$
    and $k_L^{\rm maos} - k_L^{\rm true}$ for the full event set
    (left panel), and top 10\% end-point  events of $M_{T2}$
    (right panel)
    with the input mass $\tilde m_\chi = m_{\tilde\chi_1^0}$ for the
    model point A.
    For the MAOS1 scheme, the input mass for the parent particle
    $\tilde m_Y$ is set to $m_{\tilde\chi_2^0}$.
  }
  \label{fig:deltaK}
\end{figure}

To see the characteristic feature of the MAOS momenta,
in Fig.~\ref{fig:deltaK} we show the
difference between the reconstructed and the true momenta
of $\tilde\chi_1^0$ for point A.
The left panel includes the distributions of the full event set
for the process (\ref{eq:heavy_higgs}) generated at the LHC
condition, while the right panel shows the distributions of the
top 10\% subset events near the end point of $M_{T2}$.
By definition, each MAOS scheme gives the same transverse MAOS
momenta.
For $k_L^{\rm maos} - k_L^{\rm true}$ in the MAOS1 and MAOS2 schemes,
we construct their distributions using all the possible solutions
in each event.
The result shows that the MAOS1 scheme is slightly better
than the others if one considers the full event set.
However, all the schemes provide a similar performance if one
employs a suitable $M_{T2}$ cut.
This observation and the fact that
the MAOS1 scheme cannot be adopted
when one or both parent particles are off shell
suggest that
the MAOS2 or MAOS3 schemes can be used safely without
a big loss of efficiency and regardless of whether
the parent particles are on shell or not.
The efficiency of the MAOS momenta will also vary as the
detail of the matter content and the decay process in the model.
It depends not only on the mass spectrum, but also on the coupling
structure of the particles involved in the cascade decay.
The mass gap of the decaying particles typically affects the size of
the momentum of the visible particles, and their particular momentum
direction might be forbidden by the helicity correlations.
These consequently give rise to the different shape of the $M_{T2}$
distribution, i.e., how populous are the events near the end point. 
In any case, as pointed out in the previous section, the accuracy of
the MAOS momenta can be controlled by the $M_{T2}$ cut. 
Although the study of the detailed model dependence regarding the
efficiency of the MAOS reconstruction is beyond our scope here, 
we stress that it is worthwhile to study before the application to the
experiments.

Next, we proceed to construct the invariant mass
of $H/A$ using the MAOS momenta
following Eq.~(\ref{eq:maosMass}).
\begin{figure}[t!]
  \begin{center}
    \epsfig{figure=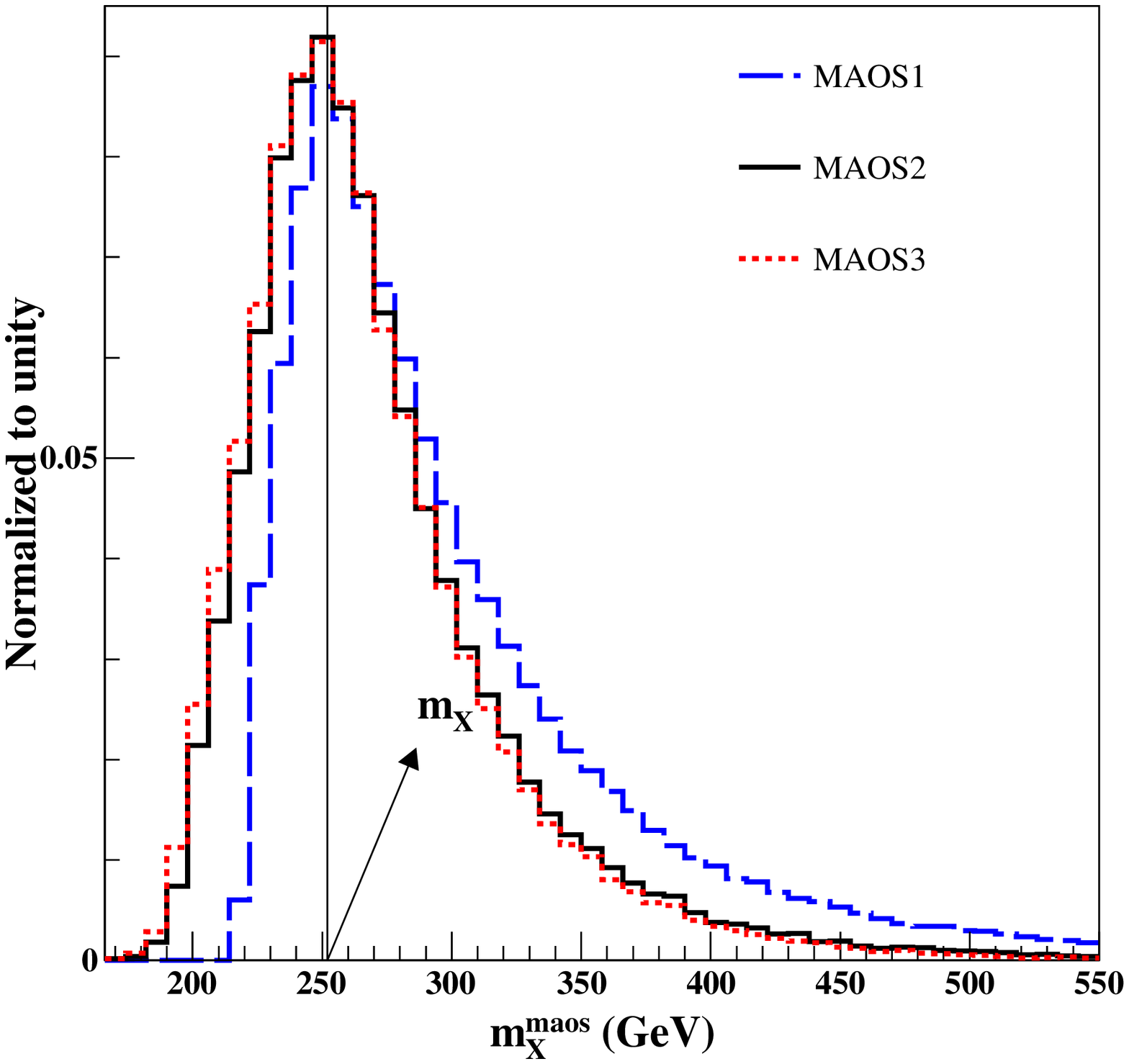,width=8cm}
    \epsfig{figure=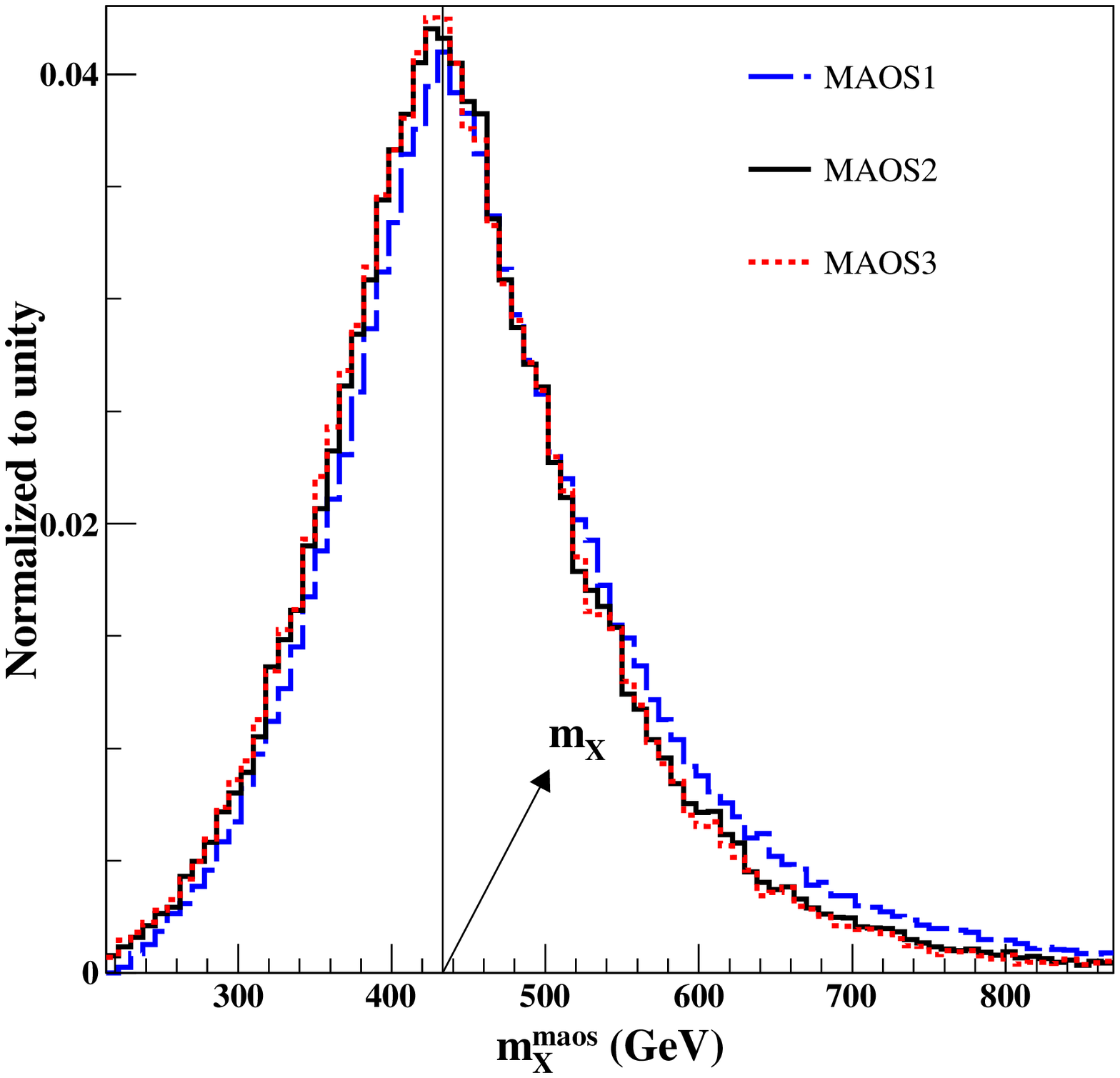,width=8cm}
   \end{center}
  \caption{The distributions of the MAOS invariant mass
    without the $M_{T2}$ cut for point A (left panel) and point B
    (right panel) with the input mass  $\tilde m_\chi =
    m_{\tilde\chi_1^0}$.
  In each panel, the MAOS1 (blue dashed), MAOS2 (black solid), and
  MAOS3 (red dotted) schemes are adopted to obtain the solution of
  the invisible momenta.
  For the MAOS1 scheme, the input mass for the parent particle mass
  $\tilde m_Y$ is set to $m_{\tilde\chi_2^0}$.
  }
  \label{fig:maosMass_schemes}
\end{figure}
In Fig.~\ref{fig:maosMass_schemes}, we show the
MAOS invariant mass distributions for the full event set
with the input trial mass $\tilde m_\chi = m_{\tilde\chi_1^0}$
while varying the MAOS scheme.
One can see that all the schemes provide a clear peak
structure around $m_{H/A}$.
This result again indicates that the performance of the
MAOS momenta does not depend much on the adopted scheme.
\begin{figure}[t!]
  \begin{center}
    {\epsfig{figure=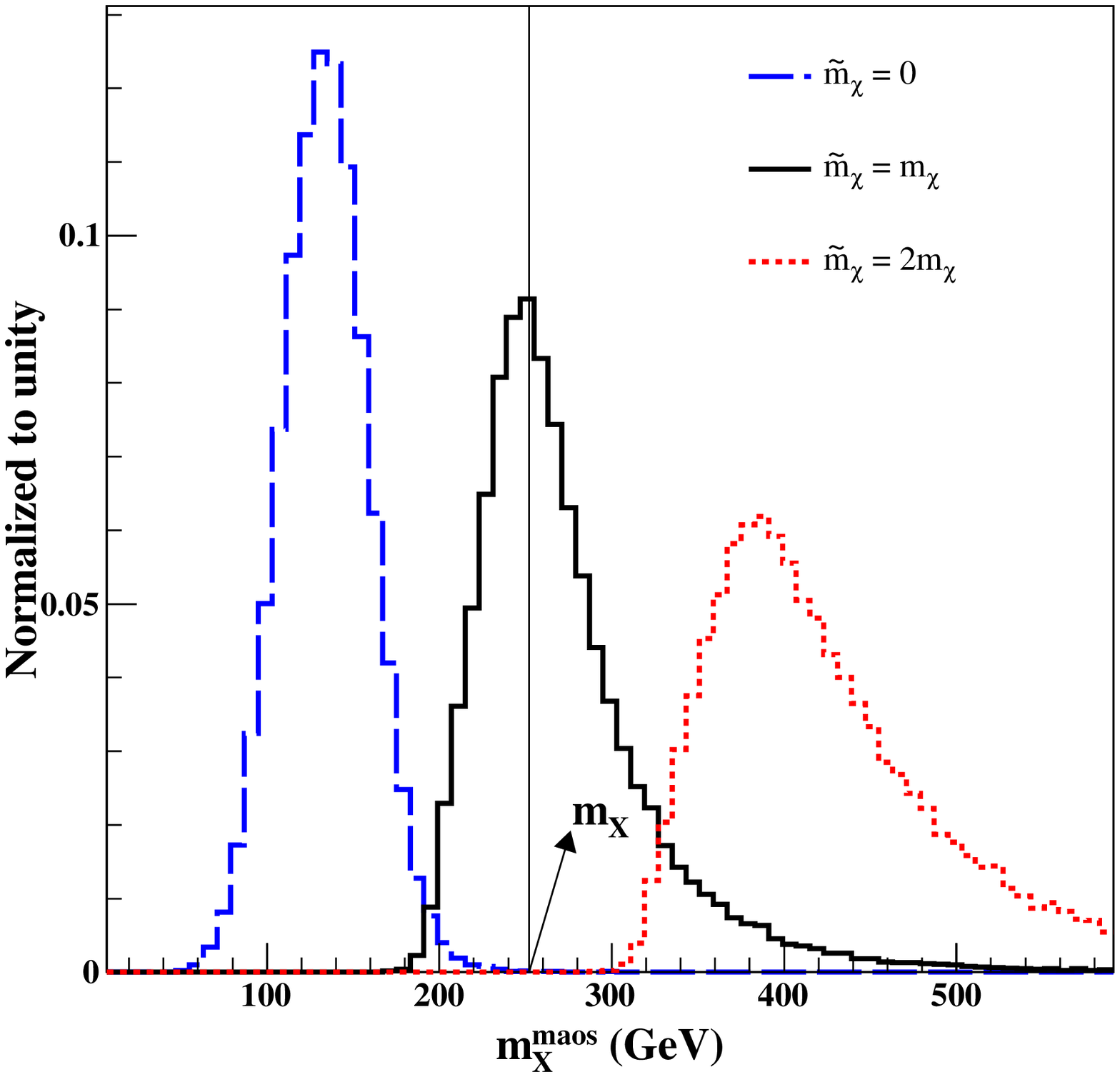,width=8cm}
     \epsfig{figure=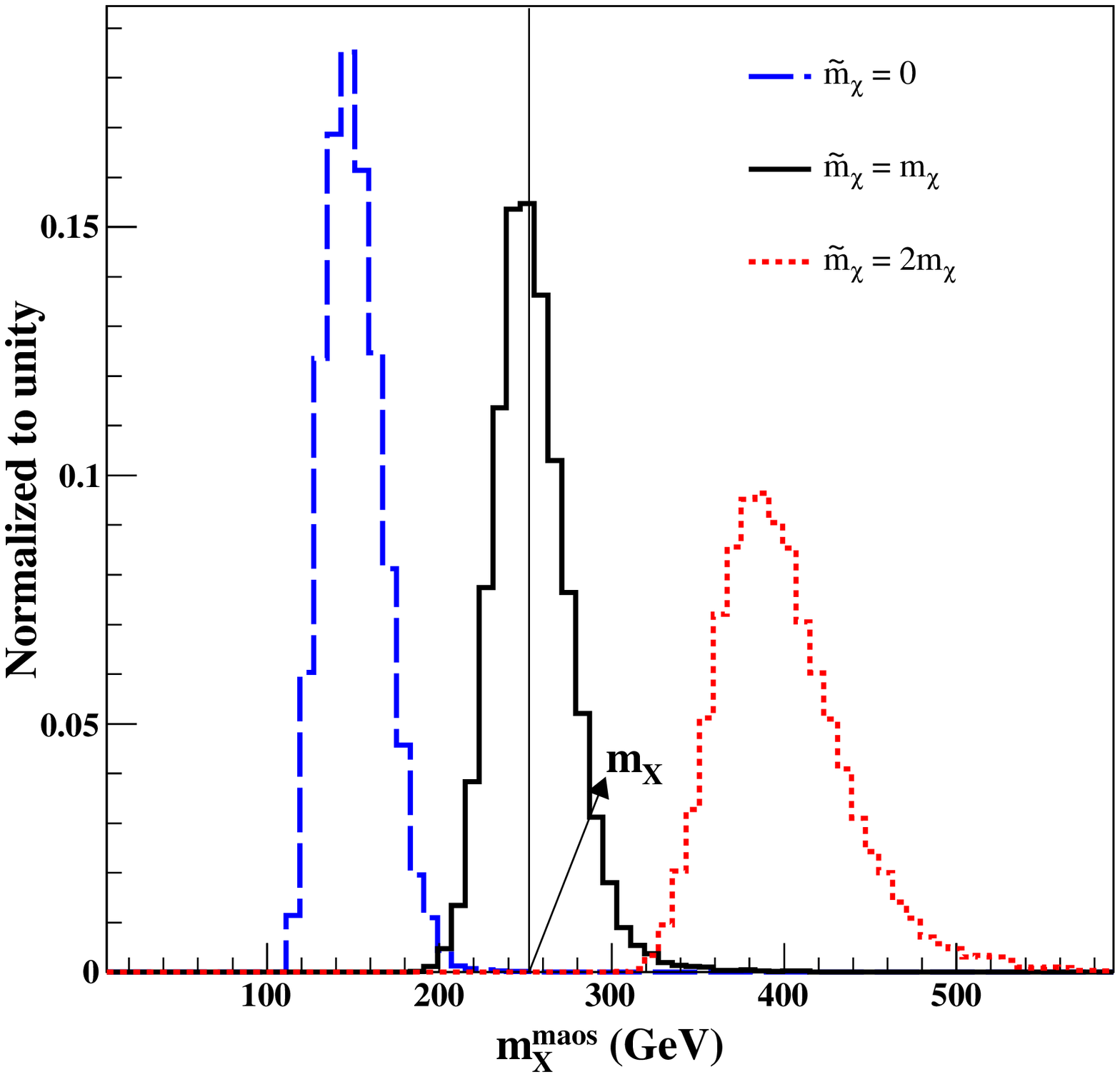,width=8cm}}
    {\epsfig{figure=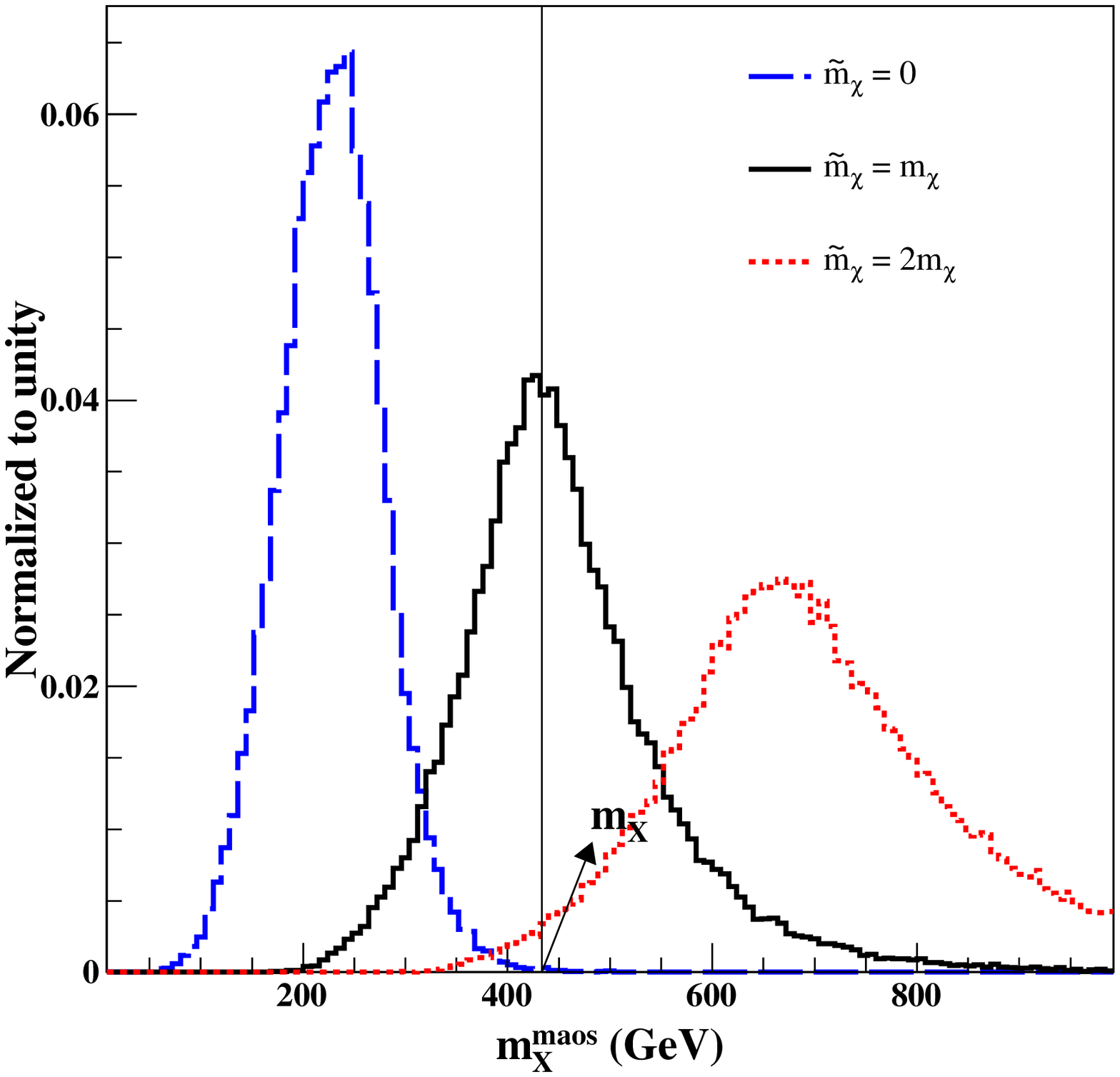,width=8cm}
     \epsfig{figure=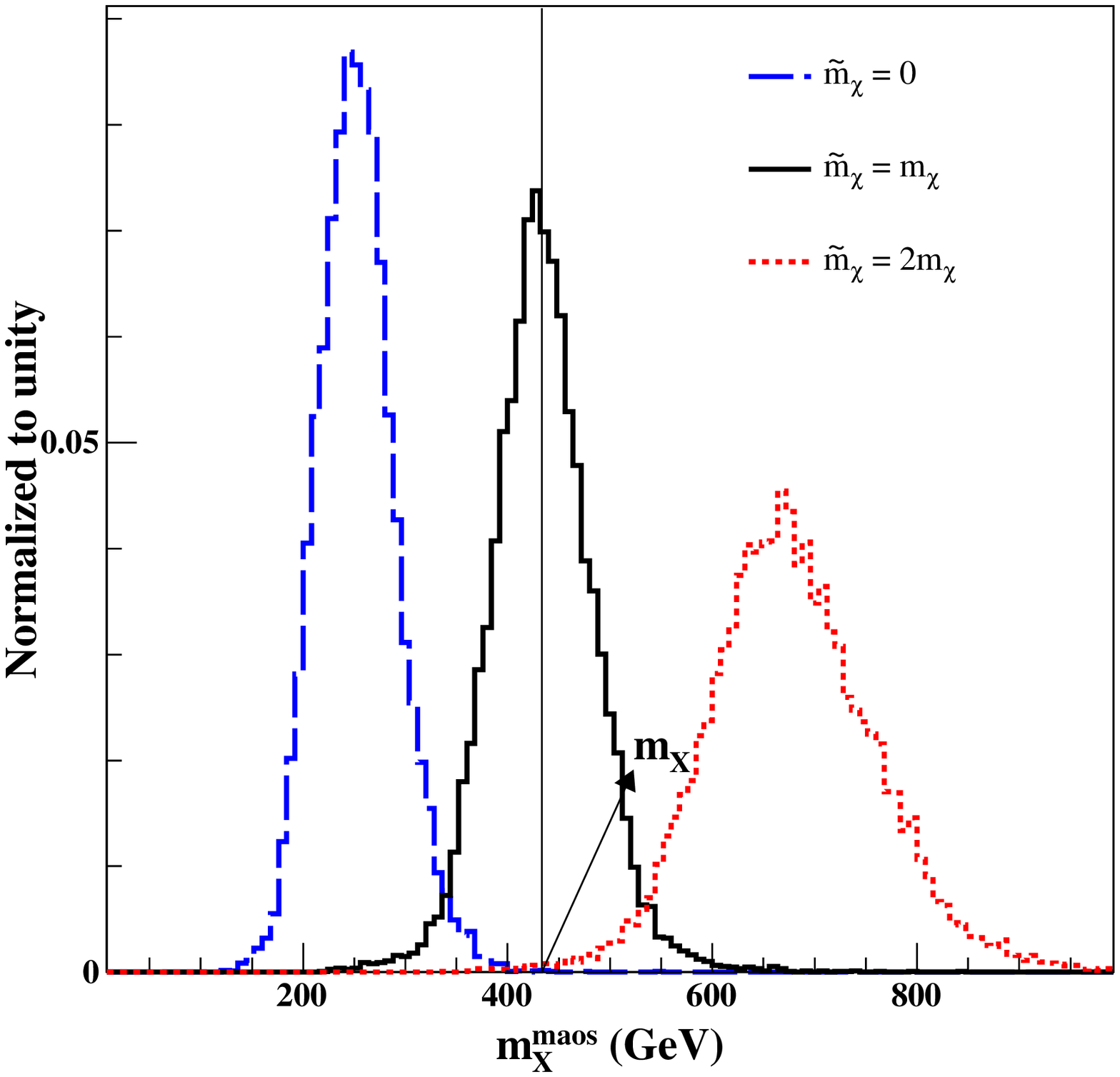,width=8cm}}
  \end{center}
  \caption{The distributions of the MAOS invariant mass for point A
  (upper panels) and point B (lower panels).
  The full event set is used in the left frames, while
  the top 30\% near-end-point events of $M_{T2}$ are used in the right
  frames. The MAOS3 scheme is adopted to obtain the solution of
  invisible momenta.}
  \label{fig:maosMass_all}
\end{figure}
Figure \ref{fig:maosMass_all} shows the dependency of the
MAOS invariant mass
on the input trial mass of the invisible particle.
Although all the distributions have a peak structure
for an arbitrary choice of the input trial mass, the
peak is located at the true resonance
mass, $m_{H/A}$, only when $\tilde m_\chi = m_{\tilde\chi_1^0}$ is
chosen.
In the right panels of Fig.~\ref{fig:maosMass_all},
the top 30\% near-end-point events of $M_{T2}$ are used.
This shows that the peak structure of
the distributions is very clear for the subset of the
events, while not changing the position of the peak.
Our simulation proves that this feature does not depend on
the MAOS scheme.

In real situations, one should consider
the combinatorial uncertainty regarding the assignment
of the visible particles to each chain.
For points A and B,
the assignment is uniquely determined
if the event includes two different pairs of opposite-sign same-flavor
leptons.
On the other hand, there are two possible combinations when
the event includes four leptons with the same flavor, i.e.
$e^+e^-e^+e^-$/$\mu^+\mu^-\mu^+\mu^-$.
One may select only the former type of events to reconstruct
the invisible momenta while sacrificing the statistics.
In a well-known model like the SM, one may calculate
the likelihood functions of kinematic variables
based on the model expectations, then select the combination
which gives the most likely solution~\cite{cdfnote}.
The other straightforward method is to use the value of
$M_{T2}$ in the event, as done in \cite{Berger:2010fy,Berger:2011ua}.
For a given event, one can construct the $M_{T2}$ of all
possible combinations, then select the combination which
gives the smallest value of the $M_{T2}$. 
It is actually
the same as the $M_{T{\rm Gen}}$ defined in \cite{Lester:2007fq}.
This method can be adopted because the $M_{T2}$ from the
correct pair will have the end point that is definitely
related to the mass spectrum of the involved particles,
whereas there is no such structure in the $M_{T2}$ from
the wrong pair, which results in a broad distribution in general.
\begin{figure}[t!]
  \begin{center}
    \epsfig{figure=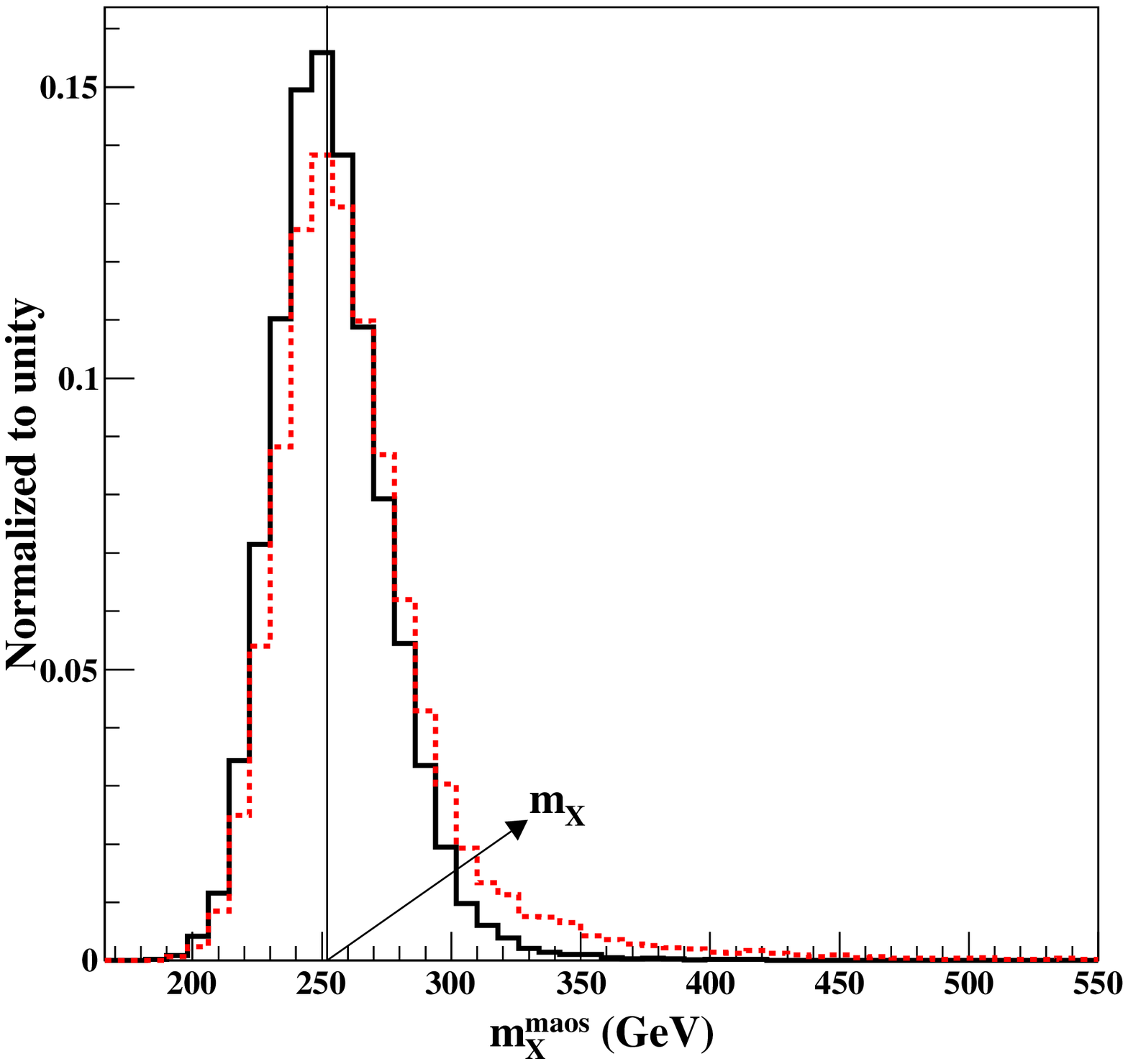,width=8cm}
    \epsfig{figure=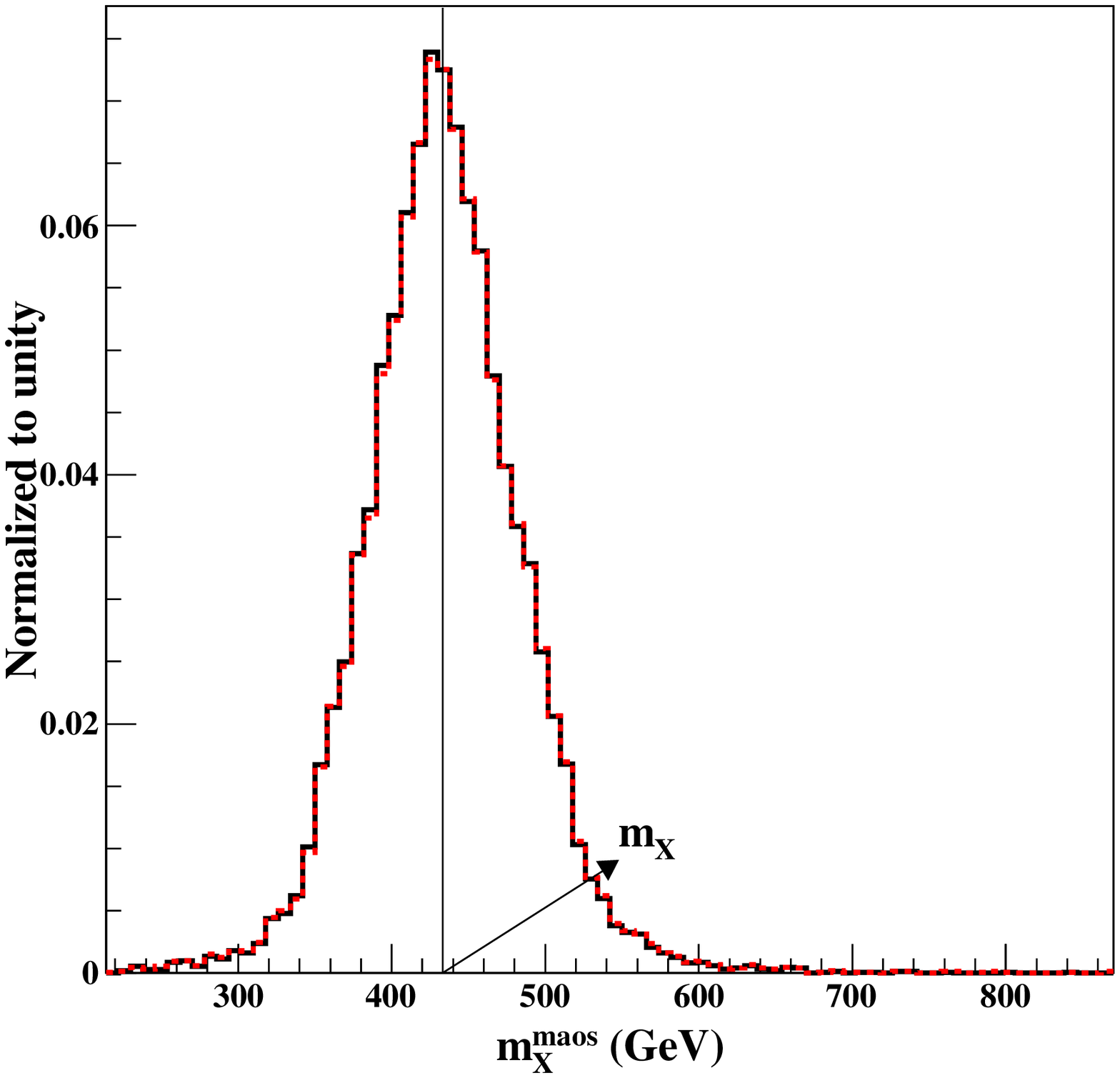,width=8cm}
  \end{center}
  \caption{The distributions of the MAOS invariant mass
    for point A (left panel) and point B  (right panel) with
    top 30\% near-end-point events of $M_{T2}$.
    The MAOS3 scheme is adopted to construct the invariant masses,
    and the input trial mass $\tilde m_\chi$ is set to $m_{\tilde\chi_1^0}$.
    In each panel, the true combination of four
    leptons is used for the black solid distribution, while
    the combination with the smallest value of $M_{T2}$ is used
    for the red dotted distribution.
  }
  \label{fig:maosMass_pairs}
\end{figure}
Our MC study shows that the method is useful for both
points A and B.
It selects the correct combination with 68\% and
97\% efficiency for points A and B, respectively.
The different level of efficiency is caused by the fact
that the decay products are relatively energetic
in point B because of heavier $H/A$, and
thus it makes the $M_{T2}$ distribution of the wrong pair
even broader.
In Fig.~\ref{fig:maosMass_pairs}, we present
the MAOS invariant mass
for the true combination (black solid), and the combination
of four leptons with the smallest value of $M_{T2}$ (red
dotted). If there is only one combination, i.e.
$e^+e^-\mu^+\mu^-$ in the event, we use the combination
in the MAOS reconstruction.
The result shows that the overall structure of the distributions
remains unchanged for both points A and B,
such that the method of finding the
correct pair using $M_{T2}$ is successful.

\section{Conclusions}
\label{sec:concl}

In this paper, we have examined the possibility of measuring
the heavy resonance mass by constructing the invariant mass
using the MAOS momenta.
Regarding the MAOS reconstruction,
we found that various schemes can be defined in order to obtain
the MAOS momenta, in particular, the longitudinal and the energy
components.
The MAOS schemes are classified by the on-shell equations as
summarized in Table \ref{table:maos_schemes}.
\begin{table}[t!]
  \caption{The definition and the number of solutions of the MAOS
    schemes.}
  \begin{center}
    \begin{tabular}{ c c cc }
      \hline\hline&&&\\[-2mm]
      & \multirow{2}*{Definition}
      & \multicolumn{2}{c}{Number of solutions}\\[2mm]
      && Events of $M_{T2}^{B}$ & Events of $M_{T2}^{UB}$
      \\[2mm]
      \hline&&&\\[-2mm]
      MAOS1 & Eqs.~(\ref{eq:on-shell_cond_chi}) and
      (\ref{eq:on-shell_cond_Y1}) & four-fold
      & four-fold\\[2mm]
      MAOS2 & Eqs.~(\ref{eq:on-shell_cond_chi}) and
      (\ref{eq:modified_scheme}) & unique
      & two-fold\\[2mm]
      MAOS3 & Eqs.~(\ref{eq:on-shell_cond_chi}) and
      (\ref{eq:unique_scheme}) & unique
      & unique\\[2mm]
      \hline\hline
    \end{tabular}
  \end{center}
  \label{table:maos_schemes}
\end{table}
Although the MAOS schemes provide different solutions in general, 
a similar level of efficiency can be obtained 
in the subset of the
events near the end point of $M_{T2}$.
We also note that the MAOS1 scheme is not applicable when
one or both parent particles $Y_i$ are off shell,
whereas the MAOS2 and MAOS3 schemes do not have such a
limitation.
Using the MAOS reconstruction of the invisible momenta,
one can construct the invariant mass of the full system.
We have shown that the peak position of the MAOS invariant mass
distribution always corresponds to the resonance mass,
at which the end point of the transverse mass distribution may
fail to point.
This feature may enable one to deduce directly the mass scale of the
heavy resonance even in the stage of early discovery,
and to measure the mass in a model-independent way.
We also expect that the MAOS invariant mass distribution
can be used as a smoking gun signal of the heavy resonance
through its clear peak structure, which is generally less
vulnerable to the background and momentum smearing effects.

\vspace{-0.2cm}
\subsection*{Acknowledgements}
\vspace{-0.3cm}
The author would like to thank K.~Choi and J.~S.~Lee for the
initial collaboration and the fruitful discussion, and also
J.~M.~Moreno and C.~Papineau for useful comments on the manuscript.
This work has been partially supported by Grants No.~FPA2010-17747,
Consolider-CPAN (CSD2007-00042) from the MICINN,
HEPHACOS-S2009/ESP1473 
from the C. A. de Madrid, and Contract 
UNILHC~PITN-GA-2009-237920 of the European Commission. \noindent


\end{document}